\documentclass[pre,twocolumn,amsmath,amssymb]{revtex4}

\usepackage{graphicx,amsmath}
\begin{document}

\title{Quantum computing via defect states in two-dimensional anti-dot lattices}

\author{Christian Flindt}
\email{cf@mic.dtu.dk}

\author{Niels Asger Mortensen}

\author{Antti-Pekka Jauho}

\affiliation{MIC -- Department of Micro and Nanotechnology,\\
NanoDTU, Technical University of Denmark,\\ Building 345east,
DK-2800 Kongens Lyngby, Denmark }

\begin{abstract}
We propose a new structure suitable for quantum computing in a
solid state environment: designed defect states in antidot
lattices superimposed on a two-dimensional electron gas at a
semiconductor heterostructure. State manipulation can be obtained
with gate control. Model calculations indicate that it is feasible
to fabricate structures whose energy level structure is robust
against thermal dephasing.
\end{abstract}

\date{December 14, 2005}

\maketitle

\vspace{1cm}

At present an intensive search is taking place for solid-state
structures which are suitable for quantum computing; a typical
example consists of  gate-defined double-dot systems studied by
several groups \cite{Loss:1998, Fujisawa:1998, Elzerman:2003,
Hayashi:2003, Gorman:2005, Johnson:2005}. A necessary requirement
for a practical application is scalability \cite{Divincenzo:2000},
and many of the existing structures do not immediately offer this
possibility. Here we propose an alternative scheme:
quantum-mechanical bound states which form at defects in an
anti-dot superlattice defined on a semiconductor heterostructure.
Scalability is not a critical issue for the suggested structures,
which enable the fabrication of a large number of solid-state
qubits with no particular extra effort. The flexibility offered by
e-beam or local oxidation techniques allows the sample designer to
optimize the samples for many different purposes with a very high
degree of control.

Anti-dot lattices on semiconductor heterostructures have been a
topic of intense research due to their interesting transport
properties.  In the semiclassical regime novel oscillatory
features in magnetoresistance have been discovered
\cite{Gerhardts:1989}, and as the lattice spacing is diminished
and the quantum regime is approached, exotic energy spectra, such
as the Hofstadter butterfly \cite{Hofstadter:1976} may become
experimentally accessible. The fabrication of anti-dot lattices
with lattice constants as small as 75 nm has been demonstrated in
experiments \cite{Stadelmann:2003}. Smaller lattice constants are
however expected to be within experimental reach
\cite{Cavallini:2003} leading to a further enhancement of quantum
effects.  We shall in this paper demonstrate that state-of-the-art
anti-dot lattices may have important practical applications in
quantum information processing.

Consider a two-dimensional electron gas (2DEG) at a GaAs
heterostructure~\cite{footnote1} superimposed with a triangular
lattice of anti-dots with lattice constant $\Lambda$. In the
effective-mass approximation the two-dimensional single-electron
Schr\"{o}dinger equation reads
\begin{equation}
\left[-\frac{\hbar^2}{2m^*}\nabla_{\mathbf{r}}^2+\sum_i
V(\mathbf{r}-\mathbf{R}_i)\right]\psi_n(\mathbf{r})=E_n\psi_n(\mathbf{r}),
\label{eq:schroedinger}
\end{equation}
where the sum runs over all anti-dots $i$, positioned at
$\mathbf{R}_i$. Each anti-dot is modelled as a circular potential
barrier of height $V_0$ and diameter $d$, \emph{i.e.}\
$V(\mathbf{r})=V_0$ for $r<d/2$, and zero elsewhere. It is
convenient to express all energies in terms of the length scale
$\Lambda$. Assuming that $V_0$ is so large that the eigenfunctions
$\psi_n$ do not penetrate into the anti-dots, \emph{i.e.}\
$\psi_n=0$ in the anti-dots, Eq.\ (\ref{eq:schroedinger})
simplifies to~\cite{footnote2}
\begin{equation}
-\Lambda^2\nabla_{\mathbf{r}}^2\psi_n(\mathbf{r})=\varepsilon_n\psi_n(\mathbf{r}),
\label{eq:simplified_schroedinger}
\end{equation}
where we have introduced the dimensionless eigenenergies
$\varepsilon_n\equiv E_n\Lambda^22m^*/\hbar^2$. For GaAs
$\hbar^2/2m^*\simeq 0.6$ $\mathrm{eVnm}^2$.

\begin{figure}[t!]
\includegraphics[height=0.35\textwidth]{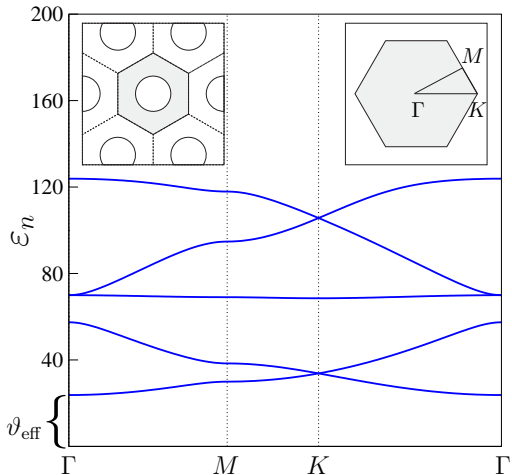}
\caption{\label{fig:bandstructure} (Color online) Bandstructure
for the periodic structure. The ratio between the diameter of the
anti-dots and the lattice constant is $d/\Lambda=0.5$. Only the
five lowest bands are shown. On the (dimensionless) energy axis we
have indicated the gap $\vartheta_{\mathrm{eff}}$ which can be
considered as the height of an effective potential (see text).
Left inset: Wigner-Seitz cell (grey area) for the periodic
structure. Circles indicate anti-dots. Right inset: First
Brillouin zone (grey area) with indications of the three
high-symmetry axes along which the bandstructure was calculated.}
\end{figure}

We first consider the perfectly periodic structure defined by the
Wigner-Seitz cell shown in the left inset of Fig.\
\ref{fig:bandstructure}. For definiteness,  we now take
$d/\Lambda=0.5$. Imposing periodic boundary conditions leaves us
with the problem of solving Eq.\
(\ref{eq:simplified_schroedinger}) on a finite-size domain. This
class of problems is well-suited for finite-element calculations,
and the available software packages make the required computations
simple, convenient, and fast \cite{Femlab}. Fig.\
\ref{fig:bandstructure} shows finite-element calculations of the
bandstructure along the high-symmetry axes indicated in the right
inset of the figure. For state-of-the-art samples $\Lambda \simeq$
75 nm, implying a band-splitting of the order of 3 meV between the
two lowest bands at the $\Gamma$-point. On the figure we have also
indicated the gap $\vartheta_{\mathrm{eff}}$ below which no states
exist for the periodic structure.

\begin{figure}[t!]
\includegraphics[height=0.4\textwidth]{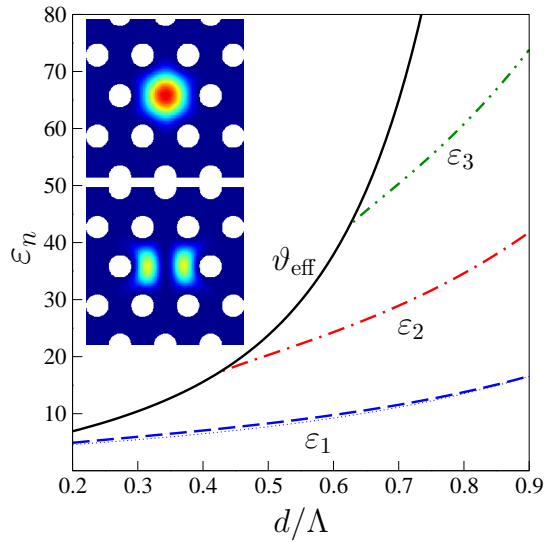}
\caption{\label{fig:singleQDspectrum} (Color online) Energy
spectrum for a single quantum dot. The three lowest dimensionless
eigenvalues, $\varepsilon_1$, $\varepsilon_2$, $\varepsilon_3$,
(corresponding to localized states) as a function of the ratio
between the anti-dot diameter $d$ and the lattice constant
$\Lambda$. The full line indicates the height
$\vartheta_{\mathrm{eff}}$ of the effective potential giving an
upper limit to the existence of bound states (see text). The thin
dotted line is the semi-analytic expression given in Eq.\
(\ref{eq:analytic_expression}). Inset: Localized eigenfunctions
$\psi_1(\mathbf{r})$ (upper panel) and $\psi_2(\mathbf{r})$
corresponding to the eigenvalues $\varepsilon_1$ and
$\varepsilon_2$, respectively, for $d/\Lambda=0.5$. The absolute
square $|\psi_i(\mathbf{r})|^2, i=1,2,$ is shown.}
\end{figure}

Next, we turn to the case where a single anti-dot has been left
out of the lattice. Relying on the analogy with photonic crystal
fibres, where similar ideas have been used to design confined
electromagnetic waves \cite{Mortensen:2005}, we expect one or
several localized states to form at the location of the `defect'.
The eigenfunctions $\psi_n$ corresponding to localized states
decay to zero within a finite distance from the defect, and it is
again sufficient to solve Eq.\ (\ref{eq:simplified_schroedinger})
on a finite-size domain. The inset in Fig.\
\ref{fig:singleQDspectrum} shows finite-element calculations of
eigenfunctions corresponding to the two lowest eigenvalues for the
geometrical ratio $d/\Lambda=0.5$. The computed energy eigenvalues
are converged with respect to an increase of the size of the
domain on which Eq.\ (\ref{eq:simplified_schroedinger}) is solved.
The two lowest eigenvalues correspond to localized states, whereas
higher eigenvalues correspond to delocalized states (not shown).
The second lowest eigenvalue is two-fold degenerate, and we only
show one of the corresponding eigenstates. One observes that the
shown eigenstate does not exhibit the underlying six-fold
rotational symmetry of the lattice. This can be traced back to the
fact that the mesh on which Eq.\
(\ref{eq:simplified_schroedinger}) was solved also lacked this
symmetry. However, as recently shown by Mortensen \emph{et al.}\
\cite{Mortensen:2004} even weak disorder in the lattice leads to a
significant deformation of the higher-order eigenstates, and the
shown eigenstate is thus likely to bear a closer resemblance to
the states occurring in  experimental structures, rather than the
one found for an ideal lattice. Similarly, we note that the
formation of defect states does not rely crucially on perfect
periodicity of the anti-dot lattice, which thus allows for a
certain tolerance in the fabrication of the anti-dot lattice.

Fig.\ \ref{fig:singleQDspectrum} also shows finite-element
calculations of the lowest eigenvalues corresponding to localized
states as a function of the geometrical ratio $d/\Lambda$. In
addition, the gap $\vartheta_{\mathrm{eff}}$ as indicated on Fig.\
\ref{fig:bandstructure} is plotted as a function of $d/\Lambda$.
The gap gives an upper limit to the existence of bounds states and
can be considered as the height of an effective two-dimensional
spherical potential well in which the localized states reside. For
GaAs with $d/\Lambda=0.5$ and $\Lambda=75$ nm the energy splitting
of the two levels is $\Delta E=E_2-E_1\simeq 1.1$ meV which is
much larger than $k_BT$ at sub-Kelvin temperatures. Thus, a
missing single anti-dot in the lattice leads to the formation of a
quantum dot with two levels at the location of the defect with an
energy level structure suitable for a charge (orbital) qubit. As
$d/\Lambda$ is increased the confinement becomes stronger and the
eigenvalues and their relative separations increase. Moreover, the
number of levels in the quantum dot can be controlled by adjusting
$d/\Lambda$, allowing for $n=1,2,3,\ldots$ levels in the quantum
dot. In particular, for any $d/\Lambda<0.42$ a single-level
quantum dot is formed.

For sample optimizing purposes it is convenient to have simple
expressions for the eigenvalues. In the limit of $d/\Lambda$
approaching 1, the problem can be approximated with that of a
two-dimensional spherical infinite potential well with radius
$\Lambda-d/2$. For this problem the lowest eigenvalue is
$\varepsilon_1^{(\infty)}=\Lambda^2\alpha_{0,1}^2/(\Lambda-d/2)^2$,
where $\alpha_{0,1}\simeq 2.405$ is the first zero of the zeroth
order Bessel function. Although this expression yields the correct
scaling with $d$, the approximation obviously breaks down for
small values of $d/\Lambda$. In that limit we follow the ideas of
Glazman \emph{et al.}\ \cite{glazman:1988} who studied quantum
conductance through narrow constrictions. The effective
one-dimensional energy barrier for transmission through two
neighboring anti-dots has a maximum value of $\pi^2$, and we thus
approximate the problem with that of a two-dimensional spherical
potential well of height $\pi^2$ and radius $\Lambda$. The lowest
eigenvalue $\varepsilon_1^{(\pi^2)}$ for this problem can be
determined numerically, and we find $\varepsilon_1^{(\pi^2)}\simeq
3.221$. Correcting for the low-$d/\Lambda$ behavior we find
\begin{equation}
\begin{split}
\varepsilon_1&\simeq\varepsilon_1^{(\infty)}-\lim_{d/\Lambda\rightarrow
0
}\varepsilon_1^{(\infty)}+\varepsilon_1^{(\pi^2)}\\&=\varepsilon_1^{(\pi^2)}+\frac{(4-d/\Lambda)d/\Lambda}{(2-d/\Lambda)^2}\alpha_{0,1}^2.
\end{split}
\label{eq:analytic_expression}
\end{equation}
In Fig.\ \ref{fig:singleQDspectrum} we show this expression
together with the results for the lowest eigenvalue determined by
finite element calculations. As can be seen on the figure, the
expression given above captures to a very high degree the results
obtained from finite-element calculations. For the higher-order
eigenvalues similar expressions can be found.

The leakage (transmission probability for penetrating the
effective potential) due to a finite size of the anti-dot lattice
can be found in the WKB approximation \cite{koshiba:2005}.
Multiplying by a characteristic attempt frequency we get the
following estimate for the inverse life-time
\begin{equation}
\frac{1}{\tau_d(E)}\sim\sqrt{\frac{E}{2m^*\Lambda^2}}
e^{-2N\Lambda \sqrt{\frac{2m^*}{\hbar^2} (V_{\mathrm{eff}}-E)}}
\label{eq:WKBapprox}
\end{equation}
where $N$ is the number of rings of anti-dots surrounding the
defect, and
$V_{\mathrm{eff}}=\vartheta_{\mathrm{eff}}\hbar^2/2m^*\Lambda^2$.
For GaAs with $\Lambda=75$ nm, $d/\Lambda=0.4$, and $N=1,2,3,4,5$,
respectively, we find $\tau_d\simeq 0.8$ ns, $0.3$ $\mu$s, $90$
$\mu$s, 30 ms, 10 s. We see that even relatively small
`superlattices' offer nearly perfect confinement.

We next consider the case where an anti-dot and one of its
next-nearest neighbors have been left out of the lattice. Due to
the close proximity of the resulting quantum dots, the different
states of the two quantum dots couple with a coupling determined
by the overlap of the corresponding single-dot wavefunctions. In
particular, for two single-level quantum dots, $L$ and $R$, with
corresponding states $|L\rangle$ and $|R\rangle$, respectively,  a
bonding $|-\rangle=(|L\rangle-|R\rangle)/\sqrt{2}$ and an
anti-bonding state $|+\rangle=(|L\rangle+|R\rangle)/\sqrt{2}$
form. The corresponding eigenenergies are $E_{\pm}=E\pm |t|$ with
$E$ being the eigenenergy corresponding to each of the states
$|L\rangle$ and $|R\rangle$, and $t$ being the tunnel matrix
element. From the eigenenergy splitting we easily obtain the
tunnel matrix element as $|t|=(E_{+}-E_{-})/2$.

\begin{figure}[t!]
\includegraphics[height=0.40\textwidth]{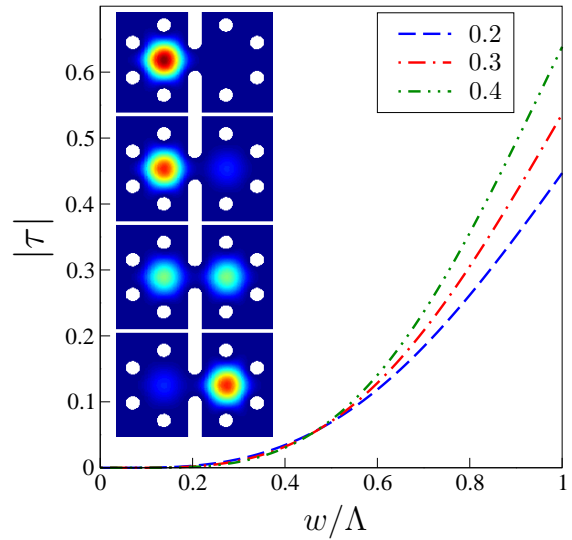}
\caption{\label{fig:coupling} (Color online) Coupling between two
single-level quantum dots. The dimensionless tunnel matrix element
$|\tau|$ as a function of the ratio between the width $w$ of the
opening defined by the split gates and the lattice constant
$\Lambda$ for different values of $d/\Lambda$ (0.2, 0.3, 0.4) in
the single-level regime. The width $w$ is defined as the shortest
distance between the split gates. Inset: Time propagation of an
electron initially prepared in the state $|L\rangle$ (uppermost
panel). Parameters are $d/\Lambda=0.4$ and $w/\Lambda=0.6$ which
for GaAs with $\Lambda=75$ nm implies as oscillation period of
$T=0.14$ ns (see text). The following panels show the state of the
electron after a time span of $T/8, 2T/8, 3T/8$ (lowest panel),
respectively. The absolute square $|\psi(\mathbf{r})|^2$ of the
electron wavefunction is shown.}
\end{figure}

The coupling of the two levels can be tuned using a metallic split
gate defined on top of the 2DEG in order to control the opening
connecting the two quantum dots. By increasing the applied gate
voltage one squeezes the opening, thereby decreasing the overlap
of the two states $|L\rangle$ and $|R\rangle$. In the following we
model the split gate with an infinite potential barrier shaped as
shown on the inset in Fig.\ \ref{fig:coupling}. Changing the
applied gate voltage effectively leads to a change of the width
$w$ of the opening, which we in the following take as a control
parameter.

In Fig.\ \ref{fig:coupling} we show finite-element calculations of
the dimensionless tunnel matrix element $|\tau| \equiv
|t|\Lambda^22m^*/\hbar^2$ as a function of the geometrical ratio
$w/\Lambda$ for a number of different values of $d/\Lambda$ in the
single-level regime, \emph{i.e.}\ $d/\Lambda<0.42$. For GaAs with
$\Lambda=75$ nm and $d/\Lambda=0.4$, $w/\Lambda=0.6$, the tunnel
matrix element is $|t|=0.015$ meV. With this coupling an electron
initially prepared in the state $|L\rangle$ is expected to
oscillate coherently between $|L\rangle$ and $|R\rangle$ with a
period of $T=h/2|t|=0.14$ ns. We note that the period agrees well
with the time scale set by the life-time obtained from  Eq.\
(\ref{eq:WKBapprox}) with $N=1$. According to the figure the
coupling varies over several orders of magnitude, thus clearly
indicating that the coupling of the two quantum dots can be
controlled via the applied gate voltage.

We have performed a numerical time propagation of an electron
initially prepared in the state $|L\rangle$. In the inset of Fig.\
\ref{fig:coupling} we show a number of snapshots at different
points in time as the electron propagates from the left to the
right quantum dot. Once located in the right quantum dot, the
electron starts propagating back to the left quantum dot (not
shown), confirming the expected oscillatory behavior.

Considering the double-dot as a charge qubit, one-qubit operations
may be performed by controlling the tunnel matrix element as
described above. Alternatively, one may consider the spin of two
electrons, each localized on one of the quantum dots, as qubits.
In that case the qubits (the spins) couple due to the exchange
coupling, which again depends on the amplitude for tunneling
between the two quantum dots. In this manner one may perform
two-qubit operations as originally proposed in Ref.
\cite{Loss:1998}.

In this work we have carried out a number of model calculations
showing that an implementation of qubits using defect states in an
anti-dot lattice is feasible. While we have here only considered
the most basic building blocks of a quantum computer, a single
charge qubit or two spin-qubits, we believe that the suggested
structure can readily be scaled to a larger number of qubits. It
is not difficult to imagine large architectures consisting of an
anti-dot lattice with several coupled defect states and/or linear
arrays of defect states constituting quantum channels along which
coherent and controllable transport of electrons can take place
\cite{Nikolopoulos:2004}. We believe that the suggested structure,
when compared to conventional gate-defined quantum dots, has the
advantage that less wiring is needed. The individual antidots need
not be electrically contacted, which in the case of conventional
gate-defined structures may be a critical issue for large
structures consisting of many quantum dots.

In conclusion, we have suggested a new structure which seems to
offer many attractive features in terms of flexibility,
scalability, and operation in the pursuit of achieving solid state
quantum computation.

The authors would like to thank D. Graf, P. E. Lindelof, and T.
Novotn\'{y} for valuable advice during the preparation of the
manuscript, and T. Markussen for sharing his numerical codes with
them.


\begin{thebibliography}{19}
\bibitem[{foo({\natexlab{a}})}]{footnote1}
\bibinfo{note}{The choice of material is not crucial for the conclusions
  reached below, and other choices of material, \emph{e.g.}\ materials with a
  lower effective mass, may be preferable.}
\bibitem[{foo({\natexlab{b}})}]{footnote2}
\bibinfo{note}{This assumption corresponds to taking the limit
  $V_0\rightarrow\infty$. We believe that our results do not change
  qualitatively for a large but finite value of $V_0$. The appealing and
  convenient form of Eq.\ (\ref{eq:simplified_schroedinger}) can however only
  be obtained in the limit $V_0\rightarrow\infty$.}
\end{thebibliography}

\end{document}